\documentclass[12pt]{article}
\usepackage{color}
\usepackage{hyperref}
\usepackage{graphicx}
\usepackage{float}
\usepackage{amsfonts}

\def\*{\vskip3mm}
\def\ie{{\it i.e\ }}\def\eg{{\it e.g.}}
\begin{document}
\large
\centerline{\large\bf About David Ruelle, after his 80th birthday}
\centerline{ Giovanni Gallavotti}
\*
\*
\begin{abstract}
This is, with minor modifications, a text read at
the 114th Statistical Mechanics meeting, in honor of D.Ruelle and Y.Sinai,
at Rutgers, Dec.13-15, 2015. It does not attempt to analyze, or not even
just quote, all works of David Ruelle; I discuss, as usual in such occasions,
a few among his works with which I have most familiarity and which were a source
of inspiration for me.
\end{abstract}

The more you read David Ruelle's works the more you are led to follow the
ideas and the references: they are, one would say, ``exciting''.  \*

The early work, on axiomatic Quantum Field Theory, today is often used and
referred to as containing the ``Haag-Ruelle'' formulation of scattering
theory, and is followed and developed in many papers, in several monographs
and in books, \cite{Ru962}. Ruelle did not pursue the subject after 1963
except in rare papers, although clear traces of the background material are
to be found in his later works, for instance the interest in the theory of
several complex variables.
%13,25,60
\*

Starting in 1963 Ruelle dedicated considerable work to Statistical
Mechanics, classical and quantum. At the time the derivation of rigorous
results and exact solutions had become of central interest, because
computer simulations had opened new horizons, but the reliability of the
results, that came out of tons of punched paper cards, needed firm
theoretical support. There had been major theoretical successes, before the
early 1950's, such as the exact solution of the Ising model and the
location of the zeros of the Ising partition function and there was a
strong revival of attention to foundations and exactly soluble models. Many
new results were about to appear; Ruelle contributed by first setting up a
proper formalism to define states via correlation functions, and then
achieving a proof of the convergence of the virial expansion. 
The result had been also obtained by C.B. Morrey in 1955, in a paper which
remained unknown in the community until the '970's, then by H.Groeneveld in
1961 and finally by O. Penrose and, independently, by Ruelle in 1963.  \*

Ruelle's contribution, \cite{Ru964}, is distinctive and original as it sets
up a clearly general method (``algebraic method'') to study the convergence
of perturbation expansions in Statistical Mechanics, Classical Mechanics,
Quantum Field Theory. Such expansions are often possible but at the cost of
expressing the objects of interest (\eg\ correlation functions or density,
magnetization, entropy, ...) as sums of power series whose $n$-th order
terms are sums of many more than the fateful $C^n$.  \*

Ruelle's systematic approach (the ``algebraic method'') to solve the
combinatorial problems has been continuously used since, giving rise to a
growing literature where results obtained via his algebraic method
approaches have been developed and merged. It made possible to many
colleagues to solve problems considered difficult in the theory of phase
transitions and in other fields. I just mention here its application to
Constructive Quantum Field theory via the the Renormalization Group.
\*

The method (today usually known as the ``cluster expansion'') is constantly
studied and improved: it contributes to a variety of fields, like to
stochastic processes, fluctuations theory, combinatorial problems. It is
remarkable that the community is divided into those who use the method and
consider it natural and many who refrain from even envisaging its use;
although this is a difficult to understand attitude, it adds to the
continued impact of Ruelle's method. The systematic theory of convergence of
a class of perturbation expansions is to be considered among the conceptually
deepest developments in the 1960's.\*

Ruelle's simultaneous more conceptual works on the foundations of
Statistical Mechanics, on the theory of the thermodynamic limit, on phase
transitions, on the proper way to address questions like ``what is a pure
state'', ``what are the conditions of stability'' for making thermodynamics
deducible from microscopic mechanics have been extremely influential: the
subject is reviewed concisely but without compromise in his book
``Statistical Mechanics'', 1969, which has become a standard part of the
curriculum of graduate students, \cite{Ru969}, and a reference book for
advanced research.  \*

A characteristic aspect of Ruelle's attitude towards science is the continuous
interplay between the need to clarify the concepts and the prodution of
unexpected solutions to concrete problems which seem to follow naturally
after the clarification.  Far from dealing only with general fundamental
questions Ruelle dealt with very concrete problems {\it among which} I mention
\\
a) the DLR equations, 1969, characterizing equilibrium states in lattice
systems, \cite{LR969}
\\
b) the theory of superstable
interactions, \cite{Ru970}, which enabled a general approach to estimates
not only in equilibrium statistical mechanics but also in quantum field
theory.
\\
c) the theory, \cite{Ru968}, of the transfer operator and its
connection with the theory of stationary states in $1D$ Statistical
Mechanics which, shortly after, plaid a surprising and unifying role in the
theory of chaotic systems.
\\
d) the first example of a phase transition in a continuous (\ie not
lattice) system, \cite{Ru971b}.
\\
e) the extension of the Lee-Yang circle theorem following Asano's work and
pushing it to deal with new cases, \cite{Ru971,Ru973b}: a work that Ruelle
kept refining and improving until quite recently,\cite{LRS012}.  \* 

An important contribution has been to establish close contact with pure
Mathematicians driving their interests to new Physics problems,
particularly in the field of Dynamical Systems. I think that the above
mentioned work, \cite{LR969}, with O.Lanford on the ``DLR equations'' has
been fundamental for the introduction into Analysis of the ``thermodynamic
formalism'' (name acquired after the title of his later, 1978, book,
\cite{Ru978}).  \*

The language adopted was often formal (particularly in the book) and not
really easy reading for a physicist but it hit the right chords in
mathematics and remains a standard reference. Many among Ruelle's studies
have strong mathematical connotation, \cite{Ru985}, and have spurred
research on subjects like the ``pressure'' of a dynamical system,
variational principles for invariant distributions, zeta functions and
periodic orbits in chaotic systems, ...  \*

The above works are among the highlights, from my limited perspective, of
the period 1963-1978.  However, starting in the early '970's, Ruelle 
introduced, with F.Takens, a new fundamental interpretation of the chaotic
motions in Fluid Mechanics, \cite{RT971}, which he shortly after developed
from a original theory of the onset of turbulence into an ambitious
theory aiming at understanding various aspects of developed turbulence: the
new theory did not meet immediate recognition perhaps because, at the time,
it was a revolution for its sharp contrast with the very foundation of
Landau's theory, namely already on the onset of turbulence.  \*

It met the fate of many novel theories of natural phenomena: many dismissed
it as ``mathematical considerations'' of little import to Physics. However
soon it became strictly interwoven with the works of M.Feigenbaum, with
several successful numerical simulations, and mainly with an ever
increasing amount of experimental evidence, starting with H.Swinney's
experiments. I still remember, at a conference, a well known experimenter
giving a talk and saying that his results were in agreement with Ruelle's
theory on the onset of turbulence, in spite of his not being able to
understand it and why.
\*

A rather early sign of the relevance of the theory is the new version of
chapter 3 in the Landau-Lifshitz book on fluids, where the onset of turbulence
based on the Ptolemaic succession of quasi periodic motions (leading to
Sec.31 of the 1959 English edition) is replaced by the sudden appearance of
a strange attractor (in the updated 1984 English version) and its
statistical properties which link the problem of developing as well as of
developed fluid turbulence to elsewhere well understood systems like 1D
Ising models with short range interaction. \*

Ruelle dedicated most of his work in the last thirty years to developing,
refining and explaining the importance of ``strange attractors'' and to
presenting and popularizing dynamical hyperbolicity as a guide to the
conceptual unification of chaotic motions and their stationary states.  \*

Besides providing tools for concrete studies, simulations and experiments
in fields apparently quite distant the unification achieved is, I think,
extremely original and deep.  \*

At this point I want to recall that Boltzmann, Clausius and Maxwell did not
hesitate to imagine microscopic motions as periodic, thus introducing and
using the often still misunderstood and vilified ergodic hypothesis to
develop equilibrium statistical mechanics. The paradigm of the
hyperbolicity (exhibited rigorously in systems like Anosov's or Axiom A
attractors) is a general paradigm, not to be dismissed (as often done) as a
mathematical fiction; rather it is a guide to turn chaotic systems into
conceptually tractable systems, by claiming their equivalence to very well
understood ones.  \*

The problem encountered by the new ideas seems to be that immediate
solutions to simple but difficult problems are expected to follow new
ideas. This means forgetting the time that has been needed to develop Gibbs
distributions into the modern theory of equilibrium: phase transitions,
phase coexistence, scaling properties in short and long range molecular
forces. The time since the late 1800's to the 1970's has been necessary to
begin (I say ``begin'') to understand equilibrium and criticality.
Similarly we have to learn how to convert the SRB distributions (which ,
in equilibrium, reduce to the now ``usual'' Gibbs distributions) into a
powerful tool to classify and understand nonequilibrium phenomena.
\*

Hyperbolic systems might turn out to play, in the modern theory of
stationary nonequilibrium, the role played by periodic motions in the early
days of Equilibrium Statistical Mechanics, with the SRB distributions
generalizing (and containing as a special case) the Gibbs distributions.
\*

Since the 1990's the focus of Ruelle's research has been on the stationary
non-equilibrium distributions of systems undergoing chaotic motions:
beginning to show the relevance for applications of the general
vision. Hence the works on strange attractors occupy a substantial fraction
of his list of publications: always paying strict attention to 
mathematical precision Ruelle has given contributions to the integral
representation of invariant measures, \cite{Ru976}, to theory of 
unstable foliations in diffeomorphisms, \cite{Ru978c}, to periodic orbits
and zeta functions, \cite{Ru976b}, to analyticity of the maximal Lyapunov
exponent in certain dynamical systems \cite{Ru979}, to ``pressure'' in
Dynamical Systems, \cite{Ru981c}, to several examples of strange
attractors, \cite{Ru981}, to application to fluid motions, \cite{Ru984}, to
new ideas and proposals on data analysis, \cite{EKR987,Ru987}, to
statistical properties of vortexes in $2D$ turbulence \cite{FR982}, to
resonances, \cite{Ru986}, to an extension of the Green-Kubo formula to
stationary states far out of equilibrium, \cite{Ru997b,Ru998b}, to
intermittency in the energy cascade, \cite{Ru014},...  \*

The works, aside from several review papers aimed at a general public, have
a formal mathematical aspect. Nevertheless they are currently being used,
for instance, in the interpretation and theory of applications to fluid
motions and atmospheric motion, \cite{Ru000b,Ru002}.  \*

In the mid '980's the state of the art on turbulence has been summarized in
a review with J.P. Eckmann which is now a standard reference, \cite{ER985}. And
Ruelle is author of several review articles and eight books, some of which
also provide a refreshing insight into the inner workings of the scientific
community.
\*

Ruelle's interest in Statistical Mechanics proper has nevertheless 
continued returning to the thermodynamic limit (\eg in spin glasses) and
inspiring also works on combinatorics, \cite{Ru999c,Ru999b,Ru010b}. The
attention to combinatorics and theory of polynomials is another facet of
his contributions intimately related and inspired by his own works on
Dynamical Systems and Statistical Mechanics.
\*

I cannot skip mentioning the close relation of Ruelle's works to the work of
Sinai: they are close in age, in methods, in mathematical clarity in defining
and studying problems from Physics, and are complementary in achievements. For
instance I think of the early contribution of Sinai on the theory of Anosov
systems, \ie on the theory of Chaos,\cite{Si968a,Si968b} and of the related
extension of the Gibbs distributions to stationary non
equilibria (the SRB distributions, named after their initials, and that of
Bowen,\cite{BR975}.\* 

Let me conclude with a personal note: I am slowly catching up to David by
age: but I remain far behind in my understanding of nature, and I know that
I am not the only one who waits to read his work and get inspired.  At
the same time I am conscious that I am falling behind in my attempts to
follow the ideas that he keeps clarifying or proposing.  I am grateful for
what I learned from him in Physics, and for his indirect influence on my
abandoning my naive mathematical formation which had addressed me in
strange directions, where the axiom of choice had a key role.
\*

\noindent{\it Acknowledgement}: I am grateful to a referee for his
intervention to clarify the original text in the form as well as in the
contents.

\small

\bibliographystyle{unsrt}
%\bibliography{0Bib}

%\begin{thebibliography}{10}

\end{document}